\def\papertitle{RAVE for Speech: Efficient Voice Conversion at High Sampling Rates}
\def\paperauthorA{Anders R. Bargum}
\def\paperauthorB{Simon Lajboschitz}
\def\paperauthorC{Cumhur Erkut}
\documentclass[twoside,a4paper]{article}
\usepackage{etoolbox}
\usepackage{dafx_24}
\usepackage{amsmath,amssymb,amsfonts,amsthm}
\usepackage{euscript}
\usepackage[T1]{fontenc}
\usepackage[utf8]{inputenc}
\usepackage{ifpdf}
\usepackage[english]{babel}
\usepackage{caption}
\usepackage{color}
\usepackage{booktabs}
\usepackage[table]{xcolor}
\usepackage{multirow}
\usepackage{subfigure}

\input glyphtounicode
\ninept

\newcounter{numauth}
\newcounter{listcnt}
\newcommand\authcnt[1]{\ifdefined#1 \stepcounter{numauth} \fi}

\newcommand\addauth[1]{
\ifdefined#1 
\stepcounter{listcnt}
\ifnum \value{listcnt}<\value{numauth}
\appto\authorslist{, #1}
\else
\appto\authorslist{~and~#1}
\fi
\fi}
\authcnt{\paperauthorB}
\authcnt{\paperauthorC}
\authcnt{\paperauthorD}
\authcnt{\paperauthorE}
\authcnt{\paperauthorF}
\authcnt{\paperauthorG}
\authcnt{\paperauthorH}
\authcnt{\paperauthorI}
\authcnt{\paperauthorJ}
\def\authorslist{\paperauthorA}
\addauth{\paperauthorB}
\addauth{\paperauthorC}
\addauth{\paperauthorD}
\addauth{\paperauthorE}
\addauth{\paperauthorF}
\addauth{\paperauthorG}
\addauth{\paperauthorH}
\addauth{\paperauthorI}
\addauth{\paperauthorJ}
\usepackage{times}
\newif\ifpdf
\ifx\pdfoutput\relax
\else
   \ifcase\pdfoutput
      \pdffalse
   \else
      \pdftrue
\fi

\ifpdf % compiling with pdflatex
  \usepackage[pdftex,
    pdftitle={\papertitle},
    pdfauthor={\authorslist},
    pdfsubject={Proceedings of the 27th International Conference on Digital Audio Effects (DAFx24)},
    colorlinks=false, % links are activated as color boxes instead of color text
    bookmarksnumbered, % use section numbers with bookmarks
    pdfstartview=XYZ % start with zoom=100% instead of full screen; especially useful if working with a big screen :-)
  ]{hyperref}
  \usepackage[pdftex]{graphicx}
\else % compiling with latex
  \usepackage[dvips]{epsfig,graphicx}
  \usepackage[dvips,
    pdftitle={\papertitle},
    pdfauthor={\authorslist},
    pdfsubject={Proceedings of the 27th International Conference on Digital Audio Effects (DAFx24)},
    colorlinks=false, % no color links
    bookmarksnumbered, % use section numbers with bookmarks
    pdfstartview=XYZ % start with zoom=100% instead of full screen
  ]{hyperref}
\fi
\usepackage[hypcap=true]{caption}
\title{\papertitle}

\threeaffiliations{
\paperauthorA \,\thanks{\vspace{-3mm}}}
{\href{https://melcph.create.aau.dk/}{Multisensory Experience Lab \& Heka VR} \\ Aalborg University\\ Copenhagen, Denmark\\
{\tt \href{mailto:arba@create.aau.dk}{arba@create.aau.dk}}
}
{\paperauthorB \,\thanks{\vspace{-3mm}}}
{\href{https://hekavr.com/}{Khora/Heka VR} \\ Copenhagen, Denmark \\ {\tt \href{mailto:simon@khora.com}{simon@khora.com}}
}
{\paperauthorC \,\thanks{\vspace{-3mm}}}
{\href{https://melcph.create.aau.dk/}{Multisensory Experience Lab} \\Aalborg University\\ Copenhagen, Denmark\\
{\tt \href{mailto:cer@create.aau.dk}{cer@create.aau.dk}}
}

\begin{document}
% more pdf-tex settings:
\ifpdf % used graphic file format for pdflatex
  \DeclareGraphicsExtensions{.png,.jpg,.pdf}
\else  % used graphic file format for latex
  \DeclareGraphicsExtensions{.eps}
\fi

%\makeatletter
%\pdfbookmark[0]{\@pdftitle}{title}
%\makeatother

\maketitle

\begin{abstract}
Voice conversion has gained increasing popularity within the field of audio manipulation and speech synthesis. Often, the main objective is to transfer the input identity to that of a target speaker without changing its linguistic content. While current work provides high-fidelity solutions they rarely focus on model simplicity, high-sampling rate environments or stream-ability. By incorporating speech representation learning into a generative timbre transfer model, traditionally created for musical purposes, we investigate the realm of voice conversion generated directly in the time domain at high sampling rates. More specifically, we guide the latent space of a baseline model towards linguistically relevant representations and condition it on external speaker information. Through objective and subjective assessments, we demonstrate that the proposed solution can attain levels of naturalness, quality, and intelligibility comparable to those of a state-of-the-art solution for seen speakers, while significantly decreasing inference time. However, despite the presence of target speaker characteristics in the converted output, the actual similarity to unseen speakers remains a challenge. 
\end{abstract}

\section{Introduction}
\label{sec:intro}
The investigation of speech synthesis, analysis, and transformation constitutes a long-standing research topic that has garnered attention for decades. Within this realm, voice conversion (VC) refers to the technique of modifying a speech signal such that it inherits specific attributes of a target speaker while keeping the linguistic content unchanged. These attributes, also referred to as features, encompass the identity, emotion, accent, or language of a target. In this study, our focus lies specifically on the conversion of speaker identity, or timbre, while maintaining linguistic and para-linguistic information. With the advent of deep learning, VC has made significant progress counting superior performance due to neural vocoders suitable for speech synthesis \cite{kong2020hifigan, kumar2019melgan}, self-supervised speech representation learning \cite{van_niekerk_comparison_2022, hsu2021hubert, qian2022contentvec} and high-fidelity feature mapping strategies \cite{choi_neural_2021, choi2023diffhiervc}. Although current models deliver relatively realistic output and exhibit high speaker adaptation, the frequent complexity of the utilized pipelines renders most models impractical in real-time scenarios. Together with the prevalent emphasis on 16kHz outputs \cite{bargum2023reimagining}, the models are therefore limited in their suitability for audio-effect applications and we are yet to experience end-to-end deep learning pipelines providing the possibility of real-time voice design for games or singing voice modifications directly in a digital audio workstation (DAW). 

In general, a main goal in VC research is zero-shot VC, also called any-to-any VC. This pertains to the ability of generalizing to unfamiliar speakers, generating spoken output that mimics the characteristics of a specific speaker, even when provided with only a brief audio sample as the target. Due to the complexity of generalising to unseen speakers without introducing mismatch problems in the different stages of the conversion pipeline, many systems are limited to any-to-many VC, meaning that any input can be converted to the speaker identities seen during training only.  

\sloppy
Few studies approach both zero-shot and any-to-many VC from a real-time perspective. In \cite{baas_Yang_2020}, a zero-shot, high-speed VC system is obtained by combining a fast neural vocoder with a decoder that maps the speech features using convolutional, linear and normalization layers only. The work in \cite{Streamableg_streamable_2022} and \cite{ning2023dualvc} similarly ensure zero-shot stream-ability by including architectures trained to operate on low-latency streaming audio such as causal-convolutions and uni-directional recurrent structures. Contrary, the work in \cite{nercRealTime} takes a signal processing-based approach, creating a singing voice conversion plugin through a source-filter method, combining pitch-shifting
with a lightweight neural filter model. 
While providing both high output quality, high fidelity and high performance, the intricacy of these models necessitates either intensive implementation efforts or dependence on external pitch shifting and formant preservation mechanisms.

In this work, we take a different approach focusing on a generative timbre transfer pipeline traditionally designed for musical purposes. Specifically, we integrate speech representation learning into the real-time variational autoencoder (RAVE) framework \cite{caillon2021rave}, enforcing linguistic content into its latent space. By treating voices as different styles we learn the ability to convert an input speech to sound like that of a target speaker at high sampling rates and low latency, both for seen and unseen targets. Our objectives encompass two primary aims: Firstly, we strive to extend the original RAVE model towards speech, preserving its initial advantages. Secondly, we endeavour to disentangle content and speaker information to explicitly control the conversions. Our proposal is grounded in the architectural design initially proposed in \cite{caillon2021rave}. We summarize our principal contributions as follows:

\sloppy
\begin{enumerate}
    \item We remove variational inference from the original RAVE model and add a feature-wise linear modulation (FiLM) conditioned auto-encoder, trained using a multi-resolution STFT loss and 3 different discriminators focusing on each their specific aspect of speech.
    \item By leveraging a speaker embedding network we inject speaker information allowing the encoder to disentangle speaker and content information.
    \item We propose information perturbation and response-based knowledge distillation techniques, employing self-supervised targets to assist the encoder in capturing content information as linguistically related soft speech units.
\end{enumerate}

\begin{figure*}[hbt!]
\center
\includegraphics[width=6.8in]{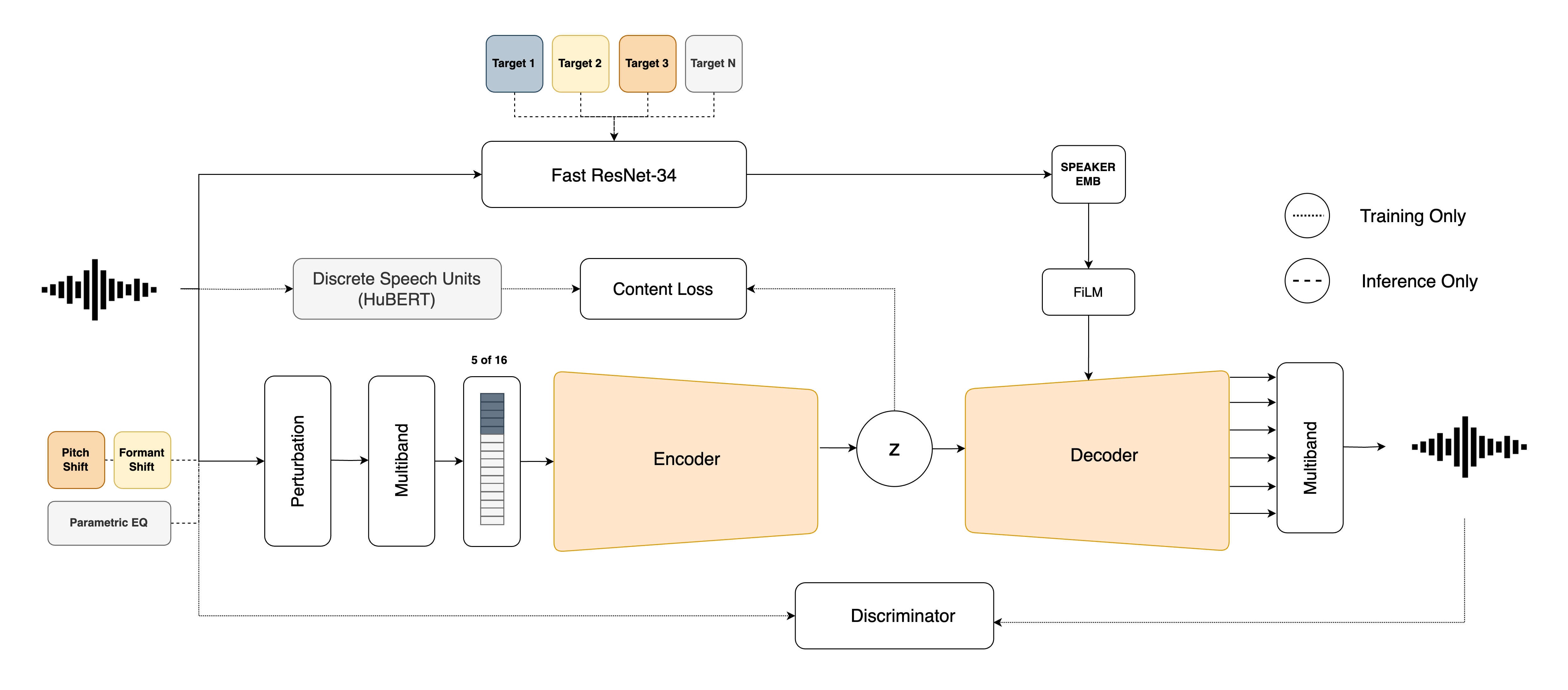}
\captionsetup{justification=centering}
\caption{\label{pipeline}{\it The proposed pipeline including our speech-related extensions to the original RAVE model. Dotted lines represent parts included exclusively while training i.e. losses and input speaker embedding, while striped lines represent aspects added for inference i.e external target audio samples. }}
\end{figure*}

\section{Related Work}

\subsection{Real-time Audio Variational autoEncoder (RAVE)}

Investigating the viability of RAVE as a foundational approach to VC presents an intriguing prospect for two primary reasons. Firstly, it has proven successful in facilitating real-time and streamable instrument transfer through the integration of multi-band decomposition and cached-causal convolutional networks. Secondly, the development of multiple exportation formats tailored to facilitate the hosting of RAVE-centric models as VST-plugins affords straightforward avenues for seamless real-time integration \cite{caillon2022streamable}\footnote{https://github.com/QosmoInc/neutone\textunderscore sdk}. Nevertheless, the variational auto-encoder-based structure employed by RAVE creates a latent space wherein all information is entangled, failing to ensure output intelligibility and maintain controllable speaker identity. Despite being trained on speech training data, the model's output has manifested as "babbling," rendering the original RAVE format unsuitable for speech synthesis and intelligible voice transfer.
\newline
\indent Two distinct works have earlier examined the feasibility of extending the RAVE model by conditioning its generation process \cite{prave, devis2023continuous}. Both of these works serve as inspiration for our approach. In \cite{devis2023continuous} they condition the RAVE model on continuous control signals, allowing for control akin to a synthesizer knob. More specifically, audio descriptors are incorporated into the generation process through fader networks providing both timbre and attribute transfer. Similarly, the work presented in \cite{prave} incorporates a stable pitch conditioning mechanism into the RAVE model to enhance the tonal signal reconstruction for singing voice synthesis. In conjunction with an ASR-based supervised phonetic encoder, these augmentations serve to improve the fidelity and controllability of the generated audio. The audio examples provided in \cite{prave} partially demonstrate intelligibility in the output.

\subsection{Zero-Shot Voice Conversion}
Feature disentanglement such as the disentanglement of speaker and content information is a fundamental principle when changing the identity of speech in a zero-shot manner i.e. when only a few utterances of a target are needed to impose its identity on the input. The incentive is to infer a speaker-independent hidden variable learned through reconstruction, which is combined with information from a new speaker when generating the converted output. In \cite{qian2019autovc}, the features are disentangled by a carefully designed bottleneck guiding the two different encoders to embed only the needed knowledge. Nonetheless, it is much more common to extract information about the incoming linguistics and speaker identity using sub-networks crafted specifically for that purpose. As for the disentanglement of linguistics, the works in \cite{kim_assem-vc_2021, nercessian20_interspeech, zhou2019modularized} utilise pre-trained content-encoders and phoneme classifiers from the field of automatic speech recognition (ASR) to extract phonetic posteriorgrams (PPGs). However, such models need large and labelled datasets and rarely generalize to languages with phonemes different to the dataset they are trained on. It is simultaneously known that PPGs may encompass speaker-related data, which could induce leakage and mismatch issues when decoding the analysis features. This will manifest as pitch fluctuations and inaccuracies when reproducing the target timbre in the generated output \cite{yang2021building}.
\newline
\indent For this reason, self-supervised speech content representations have grown popular. The authors in \cite{choi2023diffhiervc} create linguistic representations using a pre-trained Wav2Vec model whose 12th layer has proven to contain linguistic detail. Similarly, in \cite{van_niekerk_comparison_2022} and \cite{baas2022gan} k-means is used to cluster the output of a self-supervised HuBERT model in order to retrieve phonetically related discrete speech units. Speaker information may similarly be extracted using structures or pre-trained models borrowed from the speech verification community \cite{qian2019autovc} 
or through self-supervised feature representations \cite{choi_neural_2021}. In general, self-supervised speech representations (S3L) are common in the encoding and feature extraction process, ensuring efficient training and generalisation to unseen input languages. However, as the self-supervised models are complex in architecture and contain millions of parameters they may be resource-heavy and often non-useful in streaming scenarios. We investigate the possibility of incorporating the benefits of self-supervised content representation into the RAVE framework, maintaining its processing speed and simplicity. 

\section{Proposed Method: RAVE for Speech}
Our proposed network architecture is outlined in Fig. \ref{pipeline}. Compared to the original work, we map an input $x$ to a latent embedding $z$ using a traditional auto-encoder structure. The choice for utilising direct auto-encoding rather than variational inference is two-fold. Firstly, conditional VAE's do not guarantee distribution matching and often suffer from over-smoothing of the conversion \cite{qian2019autovc}. For this reason, we have seen a tendency in the VC community towards using simpler conditional AE's supported by carefully designed bottlenecks \cite{bargum2023reimagining}. Secondly, it is bad practice to mix variational and non-variational pipelines. As we aim to include a pre-trained non-variational speaker encoder, we thus opt to remove the variational inference of the original RAVE pipeline.
\newline
\indent We utilise two main advantages of the original RAVE model. Firstly, we keep the 16-band Pseudo Quadrature Mirror Filterbank (PQMF) decomposition, enabling processing at a reduced fraction of the audio system's sampling rate. Additionally, we keep the dual waveform-sub-networks enabling direct generation of natural audio from the analysis features. We anticipate that generating speech directly in the time domain will alleviate potential mismatch issues typically occurring between intermediate acoustic representations and neural vocoders in traditional VC pipelines. To enforce linguistic content in the latent embedding we utilise response-based knowledge distillation based on a self-supervised HuBERT model and further condition the RAVE decoder on speaker information using FiLM conditioning. Finally, we apply adversarial training in order to improve the naturalness of the synthesised output.

\subsection{Content Extraction}
Our content encoder is a smaller version of the architecture proposed in \cite{caillon2021rave}, being a convolutional neural network with leaky ReLU activation and batch normalization. The latent space is 64 dimensional and sampled at approximately 47Hz. As earlier mentioned, the latent space of the original framework is fully entangled, implying that it does not guarantee the separation of linguistic content. Inspired by the work in \cite{van_niekerk_comparison_2022} we therefore utilise response-based knowledge distillation, guiding the encoder output $z$ towards matching targets produced by a pre-trained HuBERT "teacher" model, which is known to produce linguistically related representations \cite{hsu2021hubert}. This approach is similarly taken in \cite{yang2024streamvc}, however, they do not perturb the input or use a pre-trained ASR speaker-encoder to inject speaker information. Rather they train their model end-to-end, at 16kHz exclusively. For the teacher model we utilise the pre-trained model described in \cite{van_niekerk_comparison_2022}, which is pre-configured to produce 100-class phonetically related discrete speech units via k-means clustering. As the teacher is trained to operate on audio data with a sampling rate of 16 kHz, we only feed the first 5 bands of the multi-band decomposition into the encoder, representing audio content up to 7.5kHz when operating at a global sampling rate of 48kHz (each output of the PQMF is down-sampled to 1.5kHz).
\newline
\indent In contrast to the methodology outlined in \cite{yang2024streamvc}, we ensure that operations on content occur within a phonetically correlated frequency range. Despite this constraint, our model retains the capability to function at high-sampling rates, thereby rendering it suitable for deployment in music production and audio workstations. As in \cite{yang2024streamvc} we add a projection layer to the output of the content encoder, transforming $z$ into $z^p$, which matches the dimensionality of the targets. We calculate the intermediate loss between the projected latent space $z^{p}$, fed through a softmax function, and the discrete speech units ${y_{dsu}}$ as follows:

\begin{equation}
 \mathcal{L}_{cont}(y_{dsu}, z^p) = CE\left(y_{dsu}, \frac{e^{z^p_i}}{\sum_{j=1}^{K} e^{z^p_j}}\right),
\end{equation}

where $CE$ denotes the cross-entropy loss function. Through this procedure, we guide the encoder to produce a content representation similar to that of the pre-trained and rather complex teacher model. The external projection layer is used as a predictor that can easily be omitted during inference.
\newline \indent
To further remove any speaker-related information captured during the encoding process, we perturb the audio input to the content encoder, meaning that we audibly manipulate the input speech. Motivated by \cite{choi_neural_2021} and \cite{xie_end--end_2022} we randomly modify the characteristics of the input audio by performing pitch shifting $ps$, formant shifting $fs$ and frequency shaping through a parametric equalizer $peq$.  We follow the same perturbation chain as proposed in \cite{choi_neural_2021}:

\begin{equation}
    \hat{x} = fs(ps(peq(x))),
\end{equation}

where $\hat{x}$ is the perturbed source speech. The parameters for each function is randomly selected for every mini-batch. The rationale behind performing information perturbation is to alter the identity of the encoder input to a degree where it becomes unrecognizable while preserving the linguistic content. Consequently, the management of speaker information is exclusively reliant on the speaker embedding.

\subsection{Speaker Injection}
We use the pre-trained Fast ResNet-34 speaker encoder from \cite{Chung_2020} to represent speaker information as speaker embedding $E_s$. The Fast ResNet-34 model contains only 1.4M parameters fitting the latency requirements of our real-time end-goal very well. In accordance with the general methodology of speech disentanglement, we condition the decoder on the retrieved speaker embedding ensuring that the audio generation is explicitly controlled by information pertaining to the specified speaker, removing the importance of the content encoder to learn any speaker-related information. Contrary to related work that concatenates $E_s$ directly to the latent vector, we condition the decoder through FiLM. More specifically we add a speaker-specific scale $\gamma$ and an offset $\beta$ to the output of each residual unit $x_r$ in the decoder:
\begin{equation}
    x_{rFiLM} = \gamma \odot x_r + \beta,
\end{equation}

where $\gamma$ and $\beta$ are computed by a linear layer that takes the conditioning as input. FiLM conditioning has proven successful in related works from both the audio-effect \cite{steinmetz2021steerable} and signal/speech compression \cite{zeghidour2021soundstream} communities. We hypothesize that the application of conditioning at various stages within the bottleneck facilitates an enhanced information flow. This enables the decoder to discern the requisite quantity and type of information necessary for effective processing. 

\begin{table*}[!htb]
\renewcommand{\arraystretch}{1.2}
\centering
\captionsetup{justification=centering}
\begin{tabular}{lcccccccc}
\toprule
\multirow{2}{*}{\textbf{Test Case}} & \multirow{2}{*}{\textbf{Model}} & \multicolumn{3}{c}{\underline{\textbf{Naturalness (DNSMOS)}}} & \multicolumn{2}{c}{\underline{\textbf{Intelligibility}}} & \underline{\textbf{Speaker Similarity}} & {\underline{\textbf{Subjective MOS}}}\\
& & SIG & BAK & OVRL & WER & CER & Resemblyzer Score & Quality \\
\midrule
\multirow{3}{*}{\textit{Unseen 2 Seen}} & AGAIN-VC \cite{chen2020againvc} & 3.36 & 3.68 & 2.92 & 40.90\% & 26.42\% & \underline{72.55}\% & 1.94 $\pm$ 0.11 \\
& DiffVC \cite{popov2022diffusionbased} & \textbf{3.55} & \textbf{4.12} & \textbf{3.30} & \underline{21.18}\% & \textbf{10.89}\% & \textbf{80.12}\% & \textbf{3.79} $\pm$ 0.12 \\
& S-RAVE \textit{(ours)} & \underline{3.49} & \underline{4.04} & \underline{3.19} & \textbf{20.21}\% & \underline{10.91}\% & 70.59\% & \underline{3.47} $\pm$ 0.13 \\
\midrule
\multirow{3}{*}{\textit{Unseen 2 Unseen}} & AGAIN-VC \cite{chen2020againvc} & 3.18 & 3.75 & 2.78 & 48.79\% & 32.34\% & \underline{69.63}\% & 1.83 $\pm$ 0.12\\
& DiffVC \cite{popov2022diffusionbased} & \textbf{3.59} & \textbf{4.15} & \textbf{3.35} & \underline{19.99}\% & \textbf{9.88}\% & \textbf{76.51}\% & \textbf{3.52} $\pm$ 0.14 \\
& S-RAVE \textit{(ours)} & \underline{3.50} & \underline{4.06} & \underline{3.21} & \textbf{18.75} \% & \underline{10.25}\% & 65.20\% & \underline{2.89} $\pm$ 0.13 \\
\midrule
\rowcolor{gray!10}
\textit{Ground Truth} & - & 3.60 & 4.08 & 3.31 & 2.94\% & 1.29\% & - & 4.77 $\pm$ 0.08\\
\bottomrule
\end{tabular}
\captionsetup{justification=centering}
\caption{\textit{Objective and subjective metrics calculated for three models across seen and unseen conversions. For the MOS score, we report the mean and the 95\% confidence interval. Bold and underlined text highlights the best and second best performing models respectively.}}
\label{tab:scores}
\end{table*}

\subsection{Discriminators}
We utilise the same multi-scale discriminator (MSD) proposed in the original work for adversarial training \cite{caillon2021rave}. However, through empirical observation, we noted that in order for the generated output to deceive the multi-scale discriminator, the produced speech exhibited highly natural characteristics while suffering from a significant lack of intelligibility. We therefore support the MSD with a multi-resolution STFT discriminator and a multi-period discriminator (MPD) combining the ideas of \cite{kong2020hifigan} and \cite{jang2021universal}. Especially the multi-resolution STFT discriminator should penalize unintelligible output, as phonetic content is highly apparent in spectrograms. The additions allow us to balance the focus of the discriminative process from which we can control both the quality and the intelligibility of the output. Additionally, we incorporate a reconstruction loss grounded in the multi-resolution STFT:

\begin{equation}
\mathcal{L}_{stft}(x, \hat{x}) = \frac{1}{N}\|\log |\operatorname{STFT}(\boldsymbol{x})|-\log |\operatorname{STFT}(\hat{\boldsymbol{x}})|\|_1,
\end{equation}
\begin{equation}
\mathcal{L}_{mstft}(x, \hat{x})=\frac{1}{M} \sum_{m=1}^M \mathcal{L}_{stft}(x_m, \hat{x}_m),
\end{equation}

where $\|\cdot\|_F$ and $\|\cdot\|_1$ denote the Frobenius and L1 norms, $\mathcal{L}_{stft}(x, \hat{x})$ denotes the STFT loss for a single resolution with $N$ elements in the magnitude and $\mathcal{L}_{mstft}(x, \hat{x})$ denotes the multi-resolution STFT loss at $M$ resolutions. The final training objective of the proposed model thus accumulates to:

\begin{equation}
    \mathcal{L}_{G_{adv}} = \lambda \mathcal{L}_{G_{msd}} + \mathcal{L}_{G_{msp}} + \mathcal{L}_{G_{mstft}}
\end{equation}
\begin{equation}
    \mathcal{L}_G = \mathcal{L}_{G_{adv}} + \mathcal{L}_{mstft} + \mathcal{L}_{cont}
\end{equation}
\begin{equation}
    \mathcal{L}_D = \lambda \mathcal{L}_{D_{msd}} + \mathcal{L}_{D_{msp}} + \mathcal{L}_{D_{mstft}},
\end{equation}
\newline
where $\mathcal{L}_{G_{adv}}$ and $\mathcal{L}_{D_{adv}}$ in eq. 6 and eq. 8 contribute to the quality of synthesized speech and are combined by the adversarial losses of the generator and discriminators respectively. $\mathcal{L}_G$ in eq. 7 is the complete generator loss consisting of the adversarial loss $\mathcal{L}_{G_{adv}}$ summed with the $\mathcal{L}_{mstft}$ reconstruction loss and CE based content loss $\mathcal{L}_{cont}$.

\section{Experiments}
Our model, denoted S-RAVE, is trained on a subset of the VCTK dataset \cite{yamagishi2019vctk}. We use 50 randomly selected speakers and resample them to 48kHz. We train the model for 1.8M iterations, approximating to 70 epochs, on a single NVIDIA A100 GPU. We train using the Adam optimizer initialised with the same hyper-parameters as the original RAVE model training. The discriminator balancing term $\lambda$ is set to 0.1. Audio examples used in the evaluation can be found on the following webpage \footnote{https://rave-for-speech-audio.github.io/}.

\subsection{Evaluation}

The proposed model is evaluated against two baseline models with different implementation strategies: a diffusion-based model \textit{DiffVC} \cite{popov2022diffusionbased} achieving state-of-the-art conversion quality, and an auto-encoder with activation guidance and adaptive instance normalization \textit{AGAIN-VC} \cite{chen2020againvc}. While the former have shown state-of-the-art results, the latter is included as it is used for baseline comparisons in various related work and thus suitable for the task at hand. All models, including our own, are either trained or pre-trained on the VCTK dataset, exclusively. Furthermore, each of their training schemes entails the same reservation of unseen speakers for testing, mirroring our own training protocol. This allows to test and compare on unseen data ensuring objectivity and fairness in evaluating performance against the models.
\newline \indent
We evaluate our model against the baselines in an unseen-to-seen speaker conversion scenario (any-to-many) and an unseen-to-unseen speaker conversion scenario (zero-shot, any-to-any) using two female and two male speakers for each evaluation \footnote{We reserve \textit{p334}, \textit{p343}, \textit{p360} and \textit{p362} as unseen speakers}. We employ a random sampling technique to select 15 utterances from each of the 4 unseen speakers. Subsequently, we engage in cross-conversion among the speakers and randomly choose 4 seen targets for the seen-conversions. In the process, each utterance is converted to one of the 4 targets yielding a total of 240 seen conversions and 240 unseen conversions.
\newline \indent
We evaluate the conversions using objective metrics along three axes: naturalness, intelligibility and speaker similarity. The assessment of naturalness is conducted through the employment of DNSMOS \cite{reddy2021dnsmos}, which serves as a proxy for subjective Mean Opinion Score (MOS) evaluations. DNSMOS comprises three distinct ratings, namely Speech Quality (SIG), Background Noise (BAK), and Overall Assessment (OVRL), to evaluate the quality of speech. We measure the speaker similarity between the target and converted speaker embeddings using Resemblyzer\footnote{https://github.com/resemble-ai/Resemblyzer} and examine intelligibility through word error rate (WER) and character error rate (CER).

\subsection{Subjective Evaluation}
The objective metrics are furthermore supported by a subjective MOS asking human subjects to rate the quality of the output produced by the models. For the subjective MOS, we present 10 convenience sampled participants and 10 participants sampled through prolific\footnote{https://www.prolific.com/participant-pool} without any reported hearing loss to different conversions (N=20). For the test, we render one utterance per unseen speaker for every seen and unseen target provided. This gives 16 conversions per model for each conversion scenario. We additionally include the original input signal as a ground truth, giving 104 samples to be rated in total. The quality of each conversion is rated on a scale from 1-5 ("very poor" to "excellent") following the MOS standard for VC. We randomly shuffle the conversions and scenarios between each participant. The test was created using the web-audio-evaluation-tool \cite{waet2015}.
\newline \indent
Finally, we report and compare the synthesis speed of the different models. We additionally perform visual inspection on both speaker and content embeddings of our proposed encoder to assess the degree of disentanglement achieved within the pipeline.

\section{Results \& Discussion}
\subsection{Objective Metrics}
Table \ref{tab:scores} shows the scores for all models across the two conversions. Overall we see that the AGAIN-VC baseline performs well on the conversion task (high speaker similarity) but lacks output quality and intelligibility, whereas our proposed model produces high-quality and intelligible output, lacking in speaker similarity. The DiffVC model performs slightly better in all scores (reported with bold text). Nevertheless, upon comparing the intelligibility and naturalness scores between our proposed model and the pre-trained DiffVC, it becomes apparent that the scores are closely aligned, with S-RAVE demonstrating better performance in WER and second best in all other aspects (reported with underlined text). It is additionally evident that the intelligibility of S-RAVE does not differ significantly for seen and unseen speakers, meaning that content and speaker information to some extent must be disentangled during the information bottleneck.
\newline \indent
Both the DiffVC and the S-RAVE models match the objectively measured naturalness of the ground truth source signal when evaluated through DNSMOS. The results conclusively demonstrate that the techniques employed to integrate content information into the RAVE latent space have yielded fruitful outcomes and that the model produces natural output. 
\newline \indent
In terms of speaker similarity, there exists a small deficiency in the proposed model. Although the converted output exhibits characteristics akin to the target (70.59\% and 65.20\%), it falls short of achieving the target quality as per contemporary standards. This is in particular heard when the model is presented with speakers not encountered during the training phase. Upon analyzing the provided examples, it becomes evident that the conversion of female input to male speaker p334 exhibits deficiencies in capturing the intended characteristics, occasionally resulting in the spill of input timbre into the rendered output. The deficiency in speaker similarity may stem from several factors. Firstly, the model's training data comprises only 50 speakers, thus impeding its ability to generalize effectively to new speakers. Secondly, the utilization of the Fast ResNet-34 speaker encoder, which possesses a smaller capacity compared to related models like the ECAPA-TDNN \cite{Desplanques_2020}, could contribute to this shortfall. Thirdly, it is possible that the latent space still retains speaker-specific information introduced by the teacher-based targets themselves. As stated by \cite{qian2022contentvec}, there may be a chance that the HuBERT model is producing speaker-related speech units. Adding information perturbation as proposed in our work will thus be less effective, as the content embeddings still will be guided towards a representation that contains some speaker information. To address this challenge in future studies, it is suggested that one could introduce perturbations to the input provided to the teacher, or incorporate more sophisticated training methodologies, such as the transfer of all input data to emulate a consistent speaker's voice. As advocated in \cite{qian2022contentvec} these approaches would further induce disentanglement in the teacher model.

\begin{figure*}[!htb]
    \centering
    \subfigure[Seen content]{\includegraphics[width=0.24\textwidth]{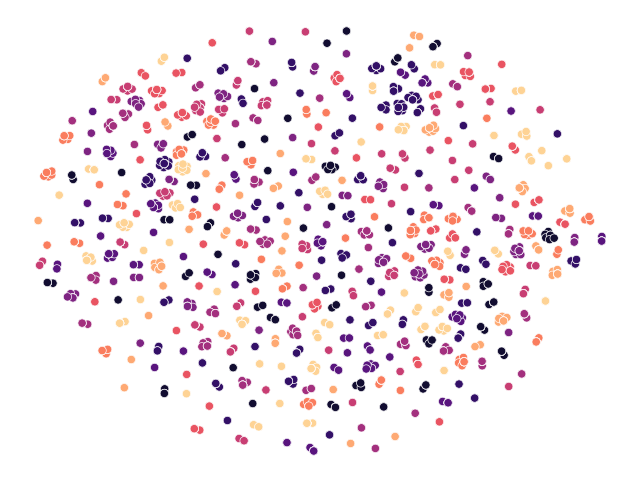}} 
    \subfigure[Seen speakers]{\includegraphics[width=0.24\textwidth]{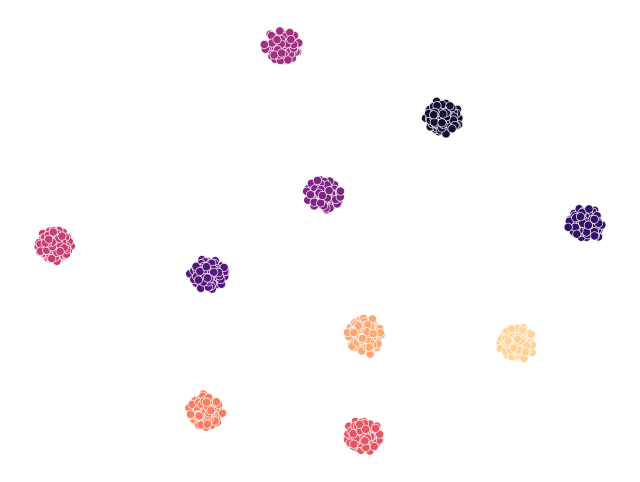}} 
    \subfigure[Unseen content]{\includegraphics[width=0.24\textwidth]{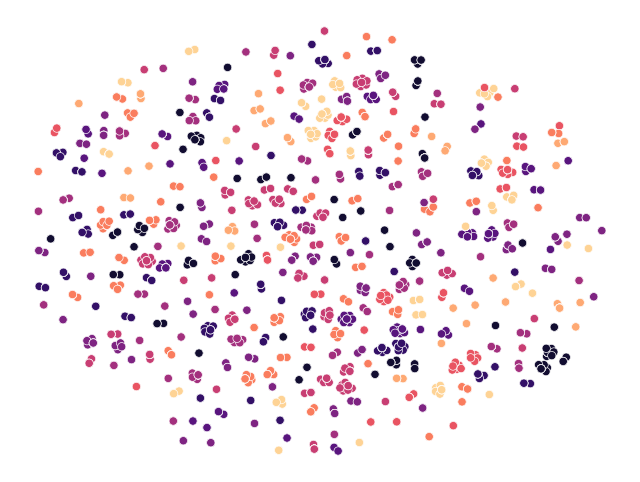}}
    \subfigure[Unseen speakers]{\includegraphics[width=0.24\textwidth]{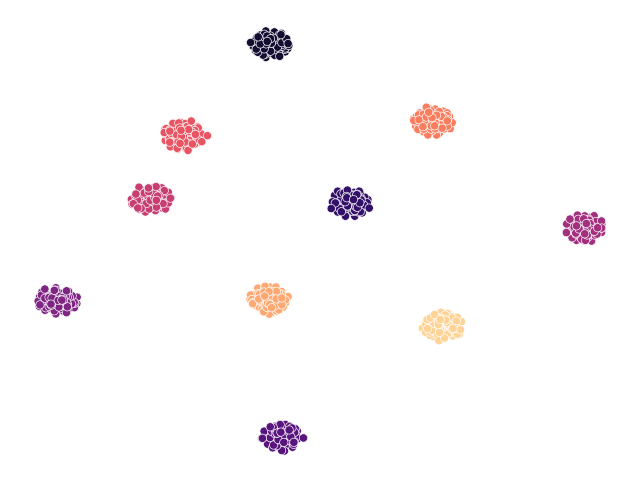}}
    \captionsetup{justification=centering}
    \caption{\textit{t-SNE visualization of the content and speaker embeddings of utterances from seen and unseen speakers. Each color represents a distinct speaker, either represented by the spoken content (a) and (c) or identity/timbre characteristics (b) and (d).}}
    \label{fig:latent}
\end{figure*}

\subsection{Subjective MOS}
The results of the subjective MOS measured on human listening subjects shows similar results as the objectively measured speech quality metrics. DiffVC performs best, closely followed by the proposed S-RAVE model, especially for the seen conversions. In contrast to the objective metrics, it is apparent that neither of the models attains the level of quality and naturalness exhibited by the ground truth. Consequently, there remains scope for enhancing the output quality of both large, intricate models like DiffVC and latency-optimized models as the proposed S-RAVE. These results furthermore indicate that the proposed model exhibits superior performance in any-to-many VC scenarios, compared with the zero-shot approach.

\subsection{Synthesis Speed}
To inspect the efficiency of the proposed model and its capability in running real-time we calculate the synthesis speed and real-time factor (RTF) of all models. We evaluate the metrics on both a CPU and an NVIDIA TITAN X (Pascal) GPU.  The synthesis speed is calculated as the average number of audio samples generated per second for 100 trials, whereas the RTF refers to the ratio of the time taken for inference to the duration of the audio signal being processed. The results, together with the amount of model parameters, are presented in table \ref{tab:speed}. 

\begin{table}[!htb]
\renewcommand{\arraystretch}{1.2}
\centering
\begin{tabular}{ccccc}
\toprule
\textbf{Model} & \textbf{Params} & & \textbf{Speed} & \textbf{RTF}\\
\midrule
\multirow{2}{*}{AGAIN-VC} & \multirow{2}{*}{7.9M} & CPU & 32.2 kHz & 0.49 \\
& & GPU & 272.55 kHz &  0.05 \\
\midrule
\multirow{2}{*}{DiffVC} & \multirow{2}{*}{126.2M} & CPU & 1.07 kHz & 20.4 \\
& & GPU & 18.97 kHz & 1.16 \\
\midrule
\multirow{2}{*}{S-RAVE (ours)} & \multirow{2}{*}{16.1M} & CPU & 706 kHz & 0.067 \\
& & GPU & 63.19 MHz & 0.007 \\
\bottomrule
\end{tabular}
\caption{\textit{Synthesis speed and RTF of the different models}}
\label{tab:speed}
\end{table}

As seen, the diffussion-based model runs rather slow both on a CPU and a GPU, which we attribute to its big generator size, diffusion process and use of external vocoder (HiFi-GAN). Contrary, AGAIN-VC runs rather fast due to its small model size. Lastly, we note that our proposed model runs around 14 times faster than real-time on a CPU and more than 100 times faster on a GPU, making it useful for real-time inference.

\subsection{Latent Embedding Visualization}
To further investigate the extent of disentanglement in the latent space, we visualize the content and speaker embeddings of 10 utterances from 10 different seen and 10 different unseen speakers. Figure \ref{fig:latent} illustrates these embeddings in 2D using Barnes-Hut t-SNE visualization. We observed that the pre-trained speaker encoder effectively groups utterances from the same speaker into distinct clusters while also separating embeddings from different speakers, regardless of whether the speakers were seen or unseen during training. Thus, the pre-trained speaker encoder successfully encapsulates speaker-specific information as intended. Conversely, we find that utterances from all speakers are dispersed across the latent space, indicating that the utterance representations are largely devoid of speaker-related information. In sub-figure (c), we note that the content embeddings from unseen speakers are more widely scattered and form larger clusters compared to those from seen speakers, suggesting that there is room for improvement in the generalizability of the content encoder. While much of the spoken content appears to be speaker-independent, certain utterances still tend to cluster according to their respective speakers. It remains uncertain whether this clustering results from phonetic similarities or the entanglement of speaker-specific information within the embedded utterances.

\section{Conclusions}
We have presented an end-to-end model for high-quality voice conversion generated directly in the time-domain at high sampling rates. The S-RAVE model, as we denote it, introduces speech representation learning techniques into an auto-encoder traditionally designed for musical timbre transfer purposes (RAVE). S-RAVE utilises information perturbation and knowledge distillation to impose content information into its latent space and uses a pre-trained speaker encoder to condition the generation process on external speaker information. Compared to the traditional RAVE model, we achieve both intelligibility and external speaker control demonstrating the effectiveness of the inserted disentanglement strategies. The model demonstrates state-of-the-art performance in converting input to seen speakers, excelling in both intelligibility and naturalness, making it sufficient for VC in a DAW or related audio-effect domains. Nevertheless, our model continues to fall short in accurately replicating the distinctive characteristics of the specified target speaker and lacks fidelity for conversions to speakers not previously encountered, as evidenced by comparisons of subjective and objective metrics with the more expansive and intricate DiffVC baseline. This means that although we incorporate the option of zero-shot conversion into the RAVE model, our proposal achieves optimal performance in any-to-many VC scenarios.

\section{Acknowledgments}
This research was partly funded by the Danish Innovation Fund. % Many thanks to the great number of anonymous reviewers. 
Additionally, we thank CLAAUDIA for access to compute hours and data services.

\nocite{*}
\bibliographystyle{IEEEbib}
\bibliography{DAFx24_tmpl} % requires file DAFx24_tmpl.bib
\end{document}